\documentclass[apj]{emulateapj}

\begin{document}
\title{Detection of a Temperature Inversion in the Broadband Infrared Emission Spectrum of TrES-4}
\author{Heather A. Knutson\altaffilmark{1,2}, David Charbonneau\altaffilmark{1,3}, Adam Burrows\altaffilmark{4}, Francis T. O'Donovan\altaffilmark{5}, and Georgi Mandushev\altaffilmark{6}}
\altaffiltext{1}{Harvard-Smithsonian Center for Astrophysics, 60 Garden St., Cambridge, MA 02138}
\altaffiltext{2}{hknutson@cfa.harvard.edu}
\altaffiltext{3}{Alfred P. Sloan Research Fellow} 
\altaffiltext{4}{Department of Astrophysical Sciences, Peyton Hall Rm. 105, Princeton, NJ 08544-1001}
\altaffiltext{5}{NASA Postdoctoral Program Fellow, Goddard Space Flight Center, 8800 Greenbelt Rd, Greenbelt, MD 20771}
\altaffiltext{6}{Lowell Observatory, 1400 West Mars Hill Rd., Flagstaff, AZ 86001}

\begin{abstract}

We estimate the strength of the bandpass-integrated thermal emission from the extrasolar planet TrES-4 at 3.6, 4.5, 5.8, and 8.0~\micron~using the Infrared Array Camera (IRAC) on the \emph{Spitzer Space Telescope}.  We find relative eclipse depths of $0.137\pm0.011\%$, $0.148\pm0.016\%$, $0.261\pm0.059\%$, and $0.318\pm0.044\%$ in these four bandpasses, respectively.  We also place a 2$\sigma$~upper limit of $0.37\%$ on the depth of the secondary eclipse in the 16~\micron~IRS peak-up array.  These eclipse depths reveal that TrES-4 has an emission spectrum similar to that of HD~209458b, which requires the presence of water emission bands created by an thermal inversion layer high in the atmosphere in order to explain the observed features.  TrES-4 receives more radiation from its star than HD 209458b and has a correspondingly higher effective temperature, therefore the presence of a temperature inversion in this planet's atmosphere lends support to the idea that inversions might be correlated with the irradiance received by the planet.  We find no evidence for any offset in the timing of the secondary eclipse, and place a $3\sigma$~upper limit of $|ecos(\omega)|<0.0058$ where $e$ is the planet's orbital eccentricity and $\omega$ is the argument of pericenter.  From this we conclude that tidal heating from ongoing orbital circulatization is unlikely to be the explanation for TrES-4's inflated radius.
                                                        
\end{abstract}


\section{Introduction}\label{intro}

Transiting extrasolar planets offer a unique opportunity to study the diversity of planetary atmospheres, and also provide an important testing ground for models of these atmospheres.  The ``hot Jupiters" are a prime example: although these planets have masses similar to the gas giant planets of the Solar system, they orbit at less than 0.05 AU from their parent stars.  Because the time scale for tidal synchronization is short compared to the ages of these systems, these planets are expected to be tidally locked, with permanent day and night sides.  This presents a significant challenge for planetary atmosphere models, as the equilibrium temperatures are substantially higher ($1000-2000$~K) and the atmospheric circulation patterns significantly different than those of Jupiter.  The hottest of these planets have temperatures comparable to those of the coolest stars, placing them in a unique parameter space with potentially exotic atmospheric chemistry.

By measuring the decrease in flux as these planets move behind their parent stars, it is possible to characterize the light emitted by the day sides of these planets, and to construct a rough spectrum \citep{char05,char08,dem05,dem06,dem07,grill07,rich07,har07,demory07,knut08,mach08}.  These observations indicate that there may be two distinct classes of hot Jupiter atmospheres.  One class of planets, including HD 189733b \citep{dem06,grill07,char08,bar08} and TrES-1 \citep{char05}, have emission spectra that are consistent with standard 1D cloud-free atmosphere models for these planets \citep{hub03,sud03,seag05,bar05,fort05,fort06a,fort06b,burr05,burr06,burr08}.  These model spectra are dominated by strong absorption features from CO and H$_2$O in the infrared.  In contrast, the emission spectra of planets such as HD 209458b \citep{dem05,rich07,burr07,knut08} and XO-1b \citep{mach08} require models with a temperature inversion between $0.1-0.01$ bars and water bands in emission instead of absorption \citep{fort06a,fort08,burr07,burr08} in order to explain the observed features.  The hot Neptune GJ 436b \citep{dem07,demory07} and the hot Saturn HD 149026b \citep{har07} have both been observed in the \emph{Spitzer} 8~\micron~bandpass, and analysis of observations at additional wavelengths is pending.

Although the nature of the high-altitude absorber needed to produce temperature inversions in the atmospheres of HD 209458b and XO-1b is currently unknown, there appears to be an intriguing connection between the equilibrium temperature of the planet in question and the presence or absence of an inversion.  HD 209458b, which has an inversion, receives two times more radiation per unit area from its primary than either HD 189733b or TrES-1, which do not have inversions.  XO-1b is an exception to this rule, as it apparently possesses a temperature inversion despite levels of irradiation comparable to those of HD 189733b and TrES-1.  Leaving XO-1b aside for the moment, it is possible that the increased irradiation experienced by HD 209458b relative to these other planets might have crossed a threshold beyond which non-equilibrium compounds form through photolysis \citep{burr08}, and that these, in turn, provide the additional opacity at altitude needed to produce the inversion.  Similarly, it has been suggested \citep{hub03,burr07,burr08,fort08} that the presence of gas-phase TiO or VO in the atmosphere might also provide the necessary opacity.  These compounds are predicted to have condensed out of the relatively cool atmospheres of HD 189733b and TrES-1, whereas they might still remain in gas phase in the hottest regions on the day side of HD 209458b.

We present below observations that allow us to test these models by examining the emission spectrum of a planet with even higher levels of irradiation than HD 209458b.  TrES-4 orbits a 1.22 M$_{\sun}$ star with a period of only 3.55 days \citep{mand07}, and as a result it receives twice as much radiation as HD 209458b, and four times as much as HD 189733b, TrES-1, and XO-1b.  Although the TrES-4 primary is more massive than the stars in these four systems it is also more distant, and has an apparent V-band magnitude of only 11.59 and a K-band magnitude of 10.33.  The planet TrES-4 is predicted to have an equilibrium temperature of approximately 1700~K, which would make it one of the hottest known transiting exoplanets.  It also has an unusually large radius, even larger than the anomalous radii of HD 209458b \citep{knut07a}, WASP-1b \citep{char07,shpo07}, and TrES-2 \citep{odon06}, which is difficult to explain without invoking additional heating processes such as ongoing tidal circularization \citep{bod01,bod03,bar03,liu08}.  Although the time scale for orbital circularization is short compared to the age of the TrES-4 system, it is possible that interactions with an unknown second planet might be pumping TrES-4's eccentricity.  The current radial velocity data set \citep{mand07}, which consists of four points, appears to be consistent with a circular orbit, although \citet{mand07} do not place an upper limit on the value of the eccentricity.  \citet{liu08} estimate that an eccentricity of approximately 0.04 could explain TrES-4's inflated radius; this would still be consistent with the radial velocity data.  By measuring the timing of the secondary eclipse of TrES-4 we will be able to constrain directly the planet's orbital eccentricity, either confirming or ruling out ongoing circularization as the explanation for the planet's inflated radius.

\section{Observations}\label{obs}

We observed three secondary eclipses of TrES-4 over a period of two weeks in Oct. 2007 using the \emph{Spitzer Space Telescope} \citep{wern04}, obtaining data at  3.6, 4.5, 5.8, 8.0, and 16 \micron.  We used the Infrared Spectrograph \citep[IRS;][]{houck04} on  UT 2007 Oct. 8 to observe an eclipse in the 16~\micron~peak-up imaging mode, acquiring a total of  736 images spanning 7.9  hours with an integration time of 30~s for each image.  Next we observed a secondary eclipse on UT 2007 Oct. 19 using the Infrared Array Camera (IRAC) \citep{faz04}, obtaining data simultaneously at 3.6 and 5.8~\micron.  Because this star is significantly dimmer than HD~189733 or HD~209458, we were able to observe in full array mode using the same 10.4 s integration time in both channels while still remaining well below saturation in the 3.6~\micron~channel, acquiring a total of 2164 images in each channel over 7.8 hours.  We observed a third secondary eclipse on UT 2007 Oct. 22 at 4.5 and 8.0~\micron, again using 10.4 s integration times in both channels and spanning 7.8 hours for a total of 2164 images in each channel.  

Because the two shortest wavelength IRAC channels (3.6 and 4.5~\micron) use InSb detectors and the three longer wavelength channels (5.8, 8.0, and 16~\micron) use Si:As detectors, there are fundamental differences between the properties of the data taken with these two types of detectors.  We describe our analysis for each type of detector separately below.

\subsection{3.6 and 4.5~\micron~Observations (InSb Detector)}\label{short_norm}

The background in the 3.6 and 4.5~\micron~IRAC bandpasses is extremely low relative to the flux from TrES-4, and contributes only 0.25\% and 0.30\%, respectively, of the total flux in an aperture with a 3-pixel radius centered on the position of the star.  As a result, we found that we obtained optimal results using aperture photometry with a radius of 3.0 pixels to estimate the flux from the star in these two channels.  We allow the position of our aperture to shift with the position of the star in each image, and see no evidence for trends correlated with the shifting position of the star on the array in the resulting time series for aperture radii larger than 2 pixels.  Our choice of a 3.0-pixel radius aperture minimizes the likelihood that transient hot pixels will be included within that aperture.  We determine the position of the star in each image as the position-weighted sum of the flux in a $5\times5$ pixel box centered on the approximate position of the star.  We estimate the background in each image from an annulus with an inner radius of 12 pixels and an outer radius of 50 pixels centered on the position of the star.  We calculate the JD value for each image as the time at mid-exposure and apply a correction to convert these JD values to the appropriate HJD, taking into account Spitzer's orbital position at each point during the observations.  

\begin{figure}
\epsscale{1.1}
\plotone{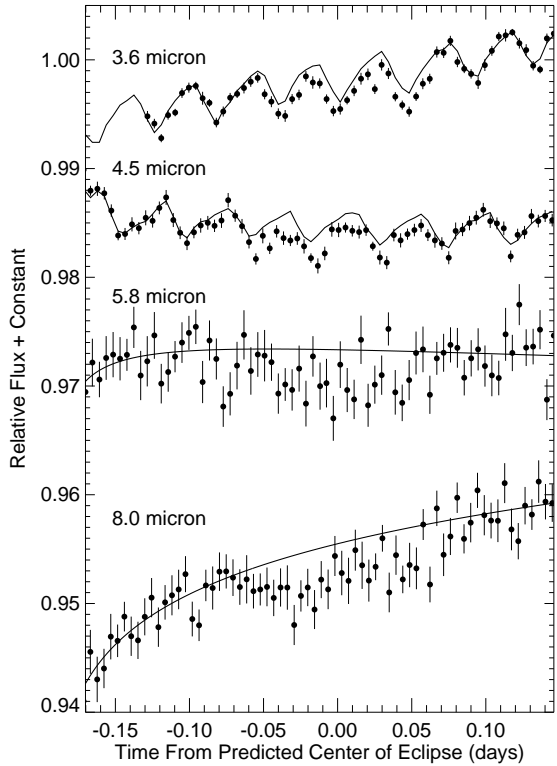}
\caption{Secondary eclipse of TrES-4 observed on UT 2007 Oct. 19 at 3.6 and 5.8~\micron~and on UT 2007 Oct. 22 at 4.5 and 8.0~\micron.  Data are binned in 6.6 minute intervals and normalized to one, then offset by a constant for the purposes of this plot.  The overplotted curves show the best-fit corrections for detector effects (see \S\ref{short_norm} and \S\ref{long_norm}). \label{norm_plots}}
\end{figure}

\begin{figure}
\epsscale{1.1}
\plotone{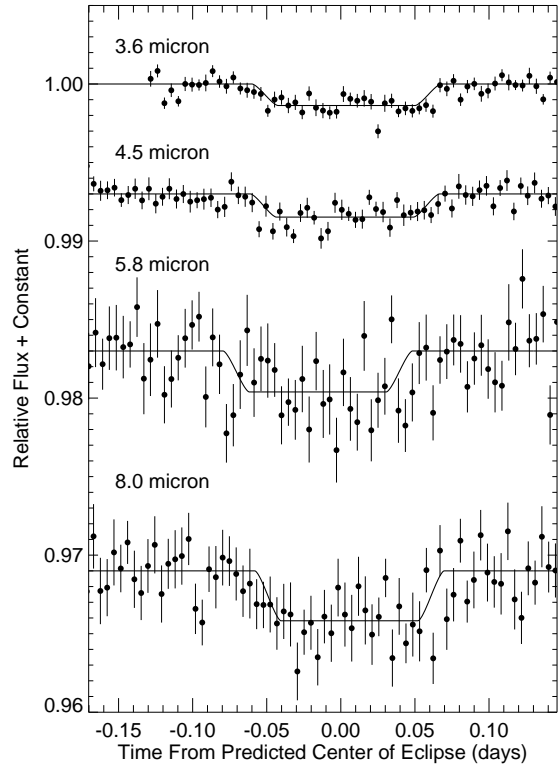}
\caption{Secondary eclipse of TrES-4 observed on UT 2007 Oct. 19 at 3.6 and 5.8~\micron~and on UT 2007 Oct. 22 at 4.5 and 8.0~\micron, with best-fit eclipse curves overplotted.  Data have been normalized to remove detector effects (see discussion in \S\ref{short_norm} and \S\ref{long_norm}), and binned in 6.6 minute intervals, then offset by a constant for the purposes of this plot.\label{four_eclipses}}
\end{figure}

Fluxes measured at these two wavelengths show a strong correlation with the changing position of the star on the array, at a level comparable to the depth of the secondary eclipse.  This effect is due to a well-documented intra-pixel sensitivity \citep{reach05,char05,char08,mor06,knut08}, and can be removed by fitting the data with a linear function of x and y position:

\begin{equation}\label{eq1}
f=f_0*(c_1+c_2(x-x_0)+c_3(y-y_0))
\end{equation}

where $f_0$ is the original flux from the star, $f$ is the measured flux, $x$ and $y$ denote the location of the flux-weighted centroid of the star on the array, $x_0$ and $y_0$ are the coordinates of the center of the pixel containing the peak of the star's point spread function, and $c_1-c_3$ are free parameters in the fit.  In the 3.6~\micron~channel $x_0$ and $y_0$ had values of [89.5,157.5], and in the 4.5~\micron~channel they had values of [84.5,156.5].  In contrast to our previous observations of HD~189733 and HD~209458 in these channels \citep{knut08,char08}, we find that adding quadratic terms to this equation does not improve the fit.  This is because the total drift in $x$ and $y$ positions during the TrES-4 observations was a factor of five smaller than for our observations of HD 209458 and HD 189733 \citep[see][for a full explanation of the pointing drifts introduced by cycling through the subarrays]{knut08}.  We find the position of the star on the array varied by 0.07 pixels in $x$ and 0.21 pixels in $y$ during our 3.6~\micron~observations.  During our 4.5~\micron~observations the position of the star varied by 0.12 pixels in $x$ and 0.14 pixels in $y$.  Because the drift in $x$ is relatively small during the 3.6~\micron~observations, we find that we obtain the same results if we fit the data with a linear function of $y$ position only, removing the linear function of $x$ from Eq. \ref{eq1}.  We use this simpler fit for the 3.6~\micron~data in order to reduce the degrees of freedom in our fit.  In the 4.5~\micron~channel we obtain optimal results using linear functions of both $x$ and $y$, as described in Eq. \ref{eq1}.  In both bandpasses the $\chi^2$ value for the fits is not improved by the addition of higher-order terms in $x$ and $y$, and the addition of these higher-order terms did not significantly alter the best-fit values for the best-fit eclipse times and depths.

After correcting for the intrapixel sensitivity, a linear trend is still visible in both channels with a slope of $+0.030\pm0.004\%$ per hour at 3.6~\micron~and $-0.020\pm0.003\%$ per hour at 4.5~\micron.  There are several possible sources of such a linear trend, including (a) variability caused by spot activity on the star, (b) the planet's phase curve, or (c) a previously uncharacterized instrumental effect related to the detector or telescope.  The total change in flux over the 8~hours spanned by these observations is larger than the secondary eclipse depth in either channel, which rules out (b) as an explanation.  TrES-4 is a late F star and thus should not have significant spot activity; the star has a rotation period $\ge 9.3$ days \citep{mand07} and our IRAC observations in these two channels were separated by only 3 days, making (b) unlikely but not impossible.  As a test we perform the same analysis on a second bright star in the array with 47\% of the flux of TrES-4.  This star shows the same positive trend at 3.6~\micron~and negative trend at 4.5~\micron~after correcting for the intrapixel sensitivity.  A preliminary analysis of similar observations of TrES-2 (F. T. O'Donovan, private commun.), which has a comparable brightness to TrES-4, also shows the same positive linear trend at 3.6~\micron~and negative linear trend at 4.5~\micron.  This argues strongly for (c) as the explanation.

We correct for this previously undocumented effect by fitting the data in both channels with a linear function of time.  This term is fitted simultaneously with the transit curve and the correction for the intrapixel sensitivity so that we can accurately characterize the additional uncertainty in the depth and timing of the eclipse introduced by these corrections.  This means that at 3.6~\micron~we are fitting for five parameters, including a constant term, a linear function of $y$ position, a linear function of time, the eclipse depth, and the eclipse time.  At 4.5~\micron~we fit for the five parameters listed above plus a linear function of $x$ position.  We also trim the first hour of data from the 3.6~\micron~time series for reasons described at the end of this section.  We fit the data using a Markov Chain Monte Carlo method \citep{ford05,winn07} with $10^6$ steps, where we set the uncertainty on individual points equal to the standard deviation of the out-of-transit data after correction for the various detector effects.  Before beginning our fits we do an intial trim to remove transient hot pixels that fell within our aperture, fitting the data with either a linear (at 3.6~\micron) or a quadratic (at 4.5~\micron) function of time to remove large trends and then discarding outliers lower than 3.5~$\sigma$ or higher than 3.0~$\sigma$.  This corresponds to an effective range of $0.992-1.007\%$ at 3.6~\micron, and a range of $0.988-1.010\%$ at 4.5~\micron.  We chose a tighter upper limit for this step because hot pixels tend to have values that are too high, rather than too low, and this limit provides more effective filtering for these pixels. 

Next we carry out the Markov chain fit on the trimmed data.  We allow both the depth and timing of the secondary eclipse to vary independently for the eclipses at each of the two observed wavelengths, and take the other parameters for the system (planetary and stellar radii, orbital period, etc.) from \citet{mand07}.  We calculate our eclipse curve using the equations from \citet{mand02} for the case with no limb-darkening.  During each step of the chain we exclude outliers greater than either 3$\sigma$~(for the 3.6~\micron~fit) or 4$\sigma$~(for the 4.5~\micron~fit), as determined using the residuals from the model light curve, from our evaluation of the $\chi^2$ function.  We rescale the value of the $\chi^2$ function to account for the fact that we are varying the number of pixels included in the fit.  Because the correction for the intra-pixel sensitivity is larger at 3.6~\micron, we find the solution in this bandpass is particularly sensitive to the presence of points with hot pixels in the time series, and we cannot obtain consistent results for the eclipse depth over a range of apertures unless we use a tight $3\sigma$ limit in our fits.  With this limit, we obtain consistent results for apertures ranging from $3-5$ pixels.  The fit in the 4.5~\micron~bandpass is less sensitive to these hot pixels, and we achieve consistent results for apertures ranging from $3-5$ pixels using a 4$\sigma$ limit.  

After running the chain, we search for the point in the chain where the $\chi^2$ value first falls below the median of all the $\chi^2$ values in the chain (i.e. where the code had first found the best-fit solution), and discard all the steps up to that point.  We take the median of the remaining distribution as our best-fit parameter, with errors calculated as the symmetric range about the median containing 68\% of the points in the distribution.  The distribution of values was very close to symmetric in all cases, and there were no strong correlations between variables.  Interestingly, we find that the slope of the linear function of time in the 3.6~\micron~fit increases with aperture size; for an aperture with a radius of 4 pixels it has increased its slope by 50\% relative to that for a 2.5 pixel aperture.  Similarly we find that the slope is steeper in the background-subtracted time series for the fainter comparison star when compared to TrES-4 photometry with the same aperture size.  Moreover, the trend in the comparison star photometry appears to be more asymptotic than linear in nature, with a steep rise of up to 1\% in the first hour and a half of observations and a more gradual slope over the rest of the time series.  This would seem to hint at an illumination-dependent effect, similar to the detector ramp described in \S\ref{long_norm} for the 5.8 and 8.0~\micron~arrays.  

\begin{figure}
\epsscale{1.2}
\plotone{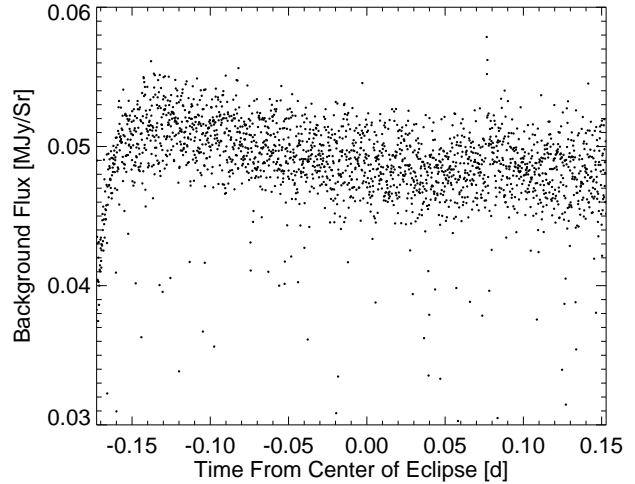}
\caption{Median background flux in the 3.6~\micron~array over the 8 hours spanned by our observations.  Note the asymptotic behavior of the measured flux values at early times.  This same asymptotic behavior is also visible in the time series for faint stars visible in the 3.6~\micron~images, albeit with a much smaller (1\% instead of 20\%) amplitude.\label{ch1_bkd}}
\end{figure}

If this effect behaves similarly to the detector ramp, we should see the same trend with a larger amplitude in the background flux, which is estimated using pixels with lower illumination levels than those of either star.  We find the median background flux in the 3.6~\micron~bandpass increases by 20\% in the first hour of observations, and then has a smaller downward trend over the rest of the observations (see Figure \ref{ch1_bkd}).  Although the background flux does not have the positive linear slope observed in the background-subtracted fluxes for TrES-4 and the comparison star, the increased amplitude (20\% vs 1\%) of the asymptotic rise in the background flux and the presence of a similar asymptote in the photometry for the comparison star, which is half the brightness of TrES-4, argues strongly for an illumination-dependent effect.  This would also provide a reasonable explanation as to why this effect was not detected in 3.6~\micron~observations of HD~189733 and HD~209458, as these stars are approximately a factor of 100 brighter than TrES-4 in this bandpass.  We elect to trim the first hour of data from the time series at 3.6~\micron~in order to remove the region that appears to exhibit asymptotic behavior.  The background in the 4.5~\micron~array has no asymptote and appears to be flat at a level of 3\% or better, therefore we continue to use the entire time series in our fits in this bandpass.

Figure \ref{norm_plots} shows the final binned data from these fits with the best fit normalizations for the detector effects in each channel overplotted, and Figure \ref{four_eclipses} shows the binned data once these trends are removed, with best-fit eclipse curves overplotted.  Best-fit eclipse depths and times are given in Table \ref{eclipse_depths}.

\subsection{5.8 and 8.0~\micron~Observations (Si:As Detector)}\label{long_norm}

At longer wavelengths the flux from the star is smaller and the zodiacal background is larger; we find that the background contributes 11\% and 12\% of the total flux in a 2.5-pixel aperture at 5.8 and 8.0~\micron, respectively.  Because the background is higher in these two channels (the median background flux is 1.43 MJy/Sr in the 5.8~\micron~bandpass and 0.71 MJy/Sr in the 8.0~\micron~bandpass), we used a psf fit to derive the time series in both bandpasses and compared the results to those from aperture photometry.  We perform aperture photometry on the images in both bandpasses using a radius of 2.5 pixels as described in \S\ref{short_norm}. For our psf fits we first calculate the background in each image by iteratively trimming $3\sigma$~outliers and fitting a Gaussian to the central region of a histogram of the remaining pixels.  We then subtract the background from the image and fit the remaining flux distribution with in-flight point spread functions generated from calibration test data\footnote{Available at http://ssc.spitzer.caltech.edu/irac/psf.html}.  We use a circular region (rounded to the nearest integer pixel) with a radius of 3.5 pixels centered on the position of the star for our psf fits, and we use the error arrays generated by the standard \emph{Spitzer} pipeline to determine the relative weighting for individual pixels.  Increasing or decreasing the radius of this region by one pixel increased the scatter in the resulting time series.  We also rescale the $x$ and $y$ coordinates of our interpolated model psf by a factor of 0.8, which effectively reduces the width of the peak of the distribution by $20\%$.  This rescaling reduces the $\chi^2$ value for our fits by a factor of two.  To fit the observed point spread function, we interpolate our model to 520 times the resolution of the IRAC array (it is already at 1/8th of the native pixel scale, and we interpolate by an additional factor of 65) and then rebin with the psf centered at the desired position, which we allow to vary in our fits.  This allows us to fit for the $x$ and $y$ position of the star to a resolution of 1/520th of a pixel.  The scatter on a representative 30-minute segment of the final fitted positions after removing a linear trend with time is $\pm0.02$~pixels, ten times larger than the minimum pixel resolution in our fits, so this is a reasonable choice.  We also fit for a constant scaling factor corresponding to the total flux. 

We flag bad pixels marked by the \emph{Spitzer} pipeline in our subarray and give them zero weight in our fits.  To find transient hot pixels, we divide the time series into sets of 20 images and calculate the median value and standard deviation at each individual pixel position within each set of 20 images.  We then step through the individual images and mark outliers more than $3\sigma$ away from the median value for that pixel position as bad pixels in that image.  We find that $98\%$ of our images have one or fewer bad pixels in the region used for our fits, which contains 29 pixels in total. This process significantly reduces the number of large outliers in the final time series, although it does not eliminate such outliers completely.  We found that increasing our threshold for bad pixels to $4\sigma$ and then $10\sigma$ outliers produced comparable results with an increasing number of large outliers in the final time series.

At 5.8~\micron~we found that the relative scatter in the time series from the psf fits was 20\% higher than in the time series from aperture photometry with a 2.5 pixel radius.  As a result of this increased scatter, which is likely produced by discrepancies between the model psf and the observed psf, we conclude that aperture photometry is still preferable in this channel.  We compare the time series using apertures ranging from $2-3.5$ pixels and find consistent results in all cases, but with a scatter that increases with the radius of the photometric aperture.  

At 8.0~\micron~we found that psf photometry produced a time series with an out-of-transit RMS variation that was 2\% lower than that of the equivalent time series using aperture photometry with a radius of 2.5 pixels.  More significantly, the use of a psf fit with bad pixel filtering effectively corrected the fluxes in 30 images with transient hot pixels in the photometric aperture, allowing them to be used in the final time series.  In light of these two changes, we conclude that psf photometry is preferable in this bandpass.

\begin{deluxetable*}{lrrrrcrrrrr}
\tabletypesize{\scriptsize}
\tablecaption{Best-Fit Eclipse Depths and Times \label{eclipse_depths}}
\tablewidth{0pt}
\tablehead{
\colhead{$\lambda$ (\micron)} & \colhead{Eclipse Depth}  & \colhead{T$_{bright}$\tablenotemark{a}} & \colhead{Center of Transit (HJD)} & \colhead{O$-$C (min.)\tablenotemark{b}}}
\startdata
3.6 & $0.137\pm0.011\%$ & $1960$\phantom{.}$\pm$\phantom{..}$70$~K & $2454392.6137\pm0.0025$ & $5.6\pm3.6$\phantom{0}\\
4.5 & $0.148\pm0.016\%$ & $1800$\phantom{.}$\pm$\phantom{..}$90$~K & $2454396.1665\pm0.0032$ & $4.0\pm4.5$\phantom{0}\\
5.8 & $0.261\pm0.059\%$ & $2210\pm300$~K & $2454392.5942\pm0.0110$ & $-22.4\pm15.8$\\
8.0 & $0.318\pm0.044\%$ & $2290\pm220$~K & $2454396.1696\pm0.0045$ & $8.4\pm6.5$\phantom{0}\\
16.0 & $<0.37\%$\tablenotemark{c}\phantom{000} & $<2350$~K\tablenotemark{c}\phantom{0} & & \\
\enddata
\tablenotetext{a}{We calculate the planet's brightness temperature in each bandpass as follows: first we set the planet's emission spectrum equal to a Planck function with the given temperature, then we take the flux-weighted average of the planet-star flux ratio over the bandpass in question, solving for the temperature required to match the observed eclipse depth in that bandpass.  We use a 6200~K Kurucz atmosphere model for the stellar spectrum; this is the same stellar spectrum used to calculate the planet-star flux ratios plotted in Fig. \ref{spectrum}.}
\tablenotetext{b}{Observed minus calculated transit times, where the expected transit times are calculated using the ephemeris from \citet{mand07} and assuming zero eccentricity.}
\tablenotetext{c}{These are the 2$\sigma$~upper limits on the eclipse depth and brightness temperature in this channel.}
\end{deluxetable*}

After determining the optimal method for estimating the fluxes in each bandpass, we must remove any detector effects in order to determine the best-fit eclipse depths.  There is no intra-pixel sensitivity in these two bandpasses, but there is another well-documented detector effect \citep{knut07b,knut08,char08} that causes the effective gain (and thus the measured flux) in individual pixels to increase over time.  This effect has been referred to as the ``detector ramp'', and has also been observed to occur in the IRS 16~\micron~peak-up array, which is made from the same material \citep{dem06}.  The size of this effect depends on the illumination level of the individual pixel.  Pixels with high ($>$250 MJy Sr$^{-1}$ in the 8~\micron~channel) will converge to a constant value within the first hour of observations, whereas lower-illumination pixels will show a linear increase in the measured flux over time with a slope that varies inversely with the logarithm of the illumination level.  In our observations of TrES-4, this effect produces a 1.5\% increase in the measured flux from the star at 8.0~\micron~during the 8 hours spanned by these observations (see Figure \ref{norm_plots}), and a much smaller (0.1\%) increase in the measured flux from the star at 5.6~\micron~over the first hour of observation.  In both bandpasses the ramp has an asymptotic shape, with a steeper rise in the first 30 minutes of observations.  We correct for this effect by fitting our time series in both bandpasses with the following quadratic function of $ln(dt)$:

\begin{equation}\label{eq2}
f=f_0*(c_1+c_2ln(dt+0.02)+c_3(ln(dt+0.02))^2)
\end{equation}

where $f_0$ is the original flux from the star, $f$ is the measured flux, and $dt$ is the elapsed time in days since the start of the observations.  In previous observations \citep{knut08,char08} we trimmed the first 30 minutes of data from the time series in order to avoid the steepest part of the ramp.  This was not necessary here, as the ramp is not as steep for fainter sources, and the addition of a constant term of 0.02 days in Eq. \ref{eq2} ensures that we are fitting the same function to these data as before.  It also ensures that we avoid the singularity at $dt=0$.  Because TrES-4 is approximately 100 times fainter than HD 209458 and HD 189733 in this bandpass, the pixels at the center of the star's psf receive lower levels of illumination and the slope of the ramp at early times is correspondingly more gradual.  Offsetting the zero point of the curve reduces the slope of the function at early times and provides a better fit to the observed behavior.  We note that our specific choice of constant has a negligible effect on the final values for the eclipse depths in these two channels; repeating these fits with a constant of 0.01 or 0.03 changes the best-fit eclipse depth at 5.8~\micron~by $0.08\sigma$ and the best-fit eclipse depth at 8.0~\micron~by $0.06\sigma$.

We fit both Eq. \ref{eq2} and the transit curve to the data simultaneously using a Markov Chain Monte Carlo method as described in \S\ref{short_norm}.  As before, the distribution of values was very close to symmetric in all cases, and there were no strong correlations between the variables.  Best-fit eclipse depths and times from these fits are given in Table \ref{eclipse_depths}, and the time series before and after correcting for detector effects are shown in Figures  \ref{norm_plots} and \ref{four_eclipses}, respectively.  As a check we repeated these fits without the quadratic term in Eq. \ref{eq2}, and found that the value of the $\chi^2$ function for our best-fit solution increased by 0.3 and 3.3 at 5.8 and 8.0~\micron, respectively.  Our choice of a linear or quadratic function has a negligible effect on the best-fit eclipse depth in the 5.8~\micron~bandpass, therefore we opt for a more consistent approach and retain the same quadratic function for our fits in both bandpasses.

\subsection{16~\micron~Observations (Si:As Detector)}\label{irs_norm}

The median background in the 16~\micron~IRS peak-up array during our observations is 1.85 MJy/Sr, more than twice the level of the background in the 8.0~\micron~bandpass, while the star is correspondingly fainter.  In this bandpass we find that the background contributes 71\% of the total flux in a 2.0 pixel aperture, and it is the primary source of noise in our final time series.  We estimate the flux from the star in each image using both aperture photometry with a radius of 2 pixels (which minimizes the RMS scatter in the resulting time series for aperture photometry) and a psf fit with a radius of 2.5 pixels, and then compare the results.  We estimate the sky background using the central $33\times44$ pixel region of the array, iteratively trimming $3\sigma$~outliers and fitting a Gaussian to the central region of a histogram of the remaining pixels.  Excluding the region containing TrES-4 from this histogram changed the median background value by only $0.02\%$, a negligible amount.  For our psf fit we calculate the median background for each image and subtract that value, then fit the remaining flux distribution with a model point spread function derived from observations of HD 42525 in this bandpass\footnote{Available at http://ssc.spitzer.caltech.edu/irs/puipsf/.}, where we allow the $x$ and $y$ position of the model to vary freely, along with a constant scaling factor corresponding to the total flux.  We use the error arrays generated by the standard \emph{Spitzer} pipeline to determine the relative weighting for individual pixels.  As described in \S\ref{long_norm},  we rescale the $x$ and $y$ coordinates of our interpolated model psf by a factor of 0.9, which effectively reduces the width of the peak of the distribution by $10\%$, in order to provide a better match for the width of the star's psf in our images.  We find that increasing or decreasing the 2.5 pixel radius of the region used for our psf fits increases the scatter in the resulting time series.  In light of the higher background fluxes in this channel, it is not surprising that we obtain better results with a smaller region than the 3.5-pixel region used in our 8.0~\micron~fits.  We find that the relative RMS scatter in the final time series is 17\% lower for psf fits than for aperture photometry, and we use the psf fits for all of our subsequent analysis.

\begin{figure}
\epsscale{1.3}
\plotone{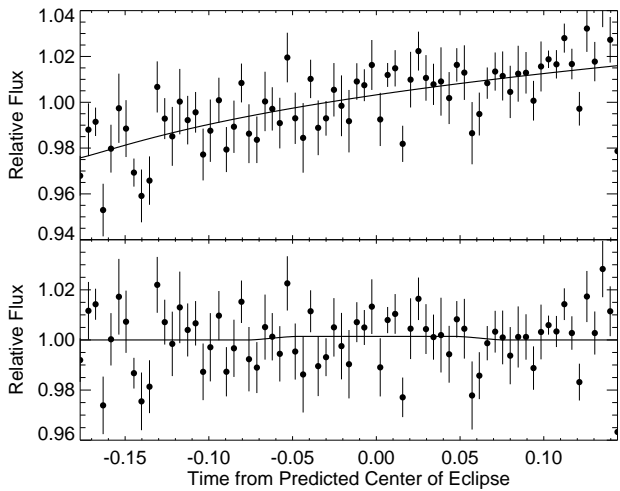}
\caption{Data from our 16~\micron~observations spanning a predicted time of secondary eclipse on UT 2007 Oct. 8.  The upper panel shows the time series with the best-fit correction for the detector ramp overplotted (see discussion in \S\ref{irs_norm}).  The lower panel shows the data after it has been normalized to remove this ramp, with the best-fit eclipse light curve overplotted.  This light curve increases slightly in flux during the eclipse event, although this increase is not statistically significant.  Data in both panels has been binned in 6.5 minute intervals.\label{irs_plot}}
\end{figure}

We plot the time series for the psf fits in the upper panel of Figure \ref{irs_plot}; as this figure illustrates, the point-to-point scatter in the time series is high enough that we are unable to detect the secondary eclipse.  We estimate the RMS variation in the time series by fitting the data outside of the predicted time of eclipse with a linear function of time, dividing by this function and then iteratively trimming 3$\sigma$ outliers.  The standard deviation of the remaining points is $3.3\%$.  We compare this to a simple estimate of the predicted noise in the 2 pixel aperture that we used for our aperture photometry, including the read noise, dark current, and photon noise from the star and background, as described in \S$7.2.3.4.4.$ of the \emph{Spitzer Observer's Manual}:

\begin{equation}\label{noise_calc}
N=\sqrt{\left(\left(i_{sky}+i_{dark}+i_{star}\right)*t_{int}+2^2\pi\frac{\left(RN\right)^2}{N_{reads}}\right)}
\end{equation}

where $N$ is the total noise in electrons, $i_{sky}$ and $i_{star}$ are the electrons generated per second in our two-pixel aperture from the sky background and the stellar flux, respectively, $i_{dark}$ is the dark current contribution, $t_{int}$ is the integration time, $RN$ is the read noise in e$^-$~s$^{-1}$ pixel$^{-1}$, and $N_{reads}$ is the number of reads.  Our images have a total integration time of 31.46 s, 16 reads per image, and we use a value of $30$~e$^-$~pixel$^{-1}$ for the read noise.  The median background flux during our observations is $1873$~e$^-$~s$^{-1}$ in our two pixel aperture, the dark current is $<126$~e$^-$~s$^{-1}$, and the star contributes a median flux of $777$~e$^-$~s$^{-1}$ in this aperture.  Evaluating Eq. \ref{noise_calc} with these numbers and assuming the dark current is negligible, we find a total noise contribution of $290$~e$^-$ in each 31.46 s exposure, which would translate to a relative RMS of $1.2\%$.  

\begin{figure}
\epsscale{1.1}
\plotone{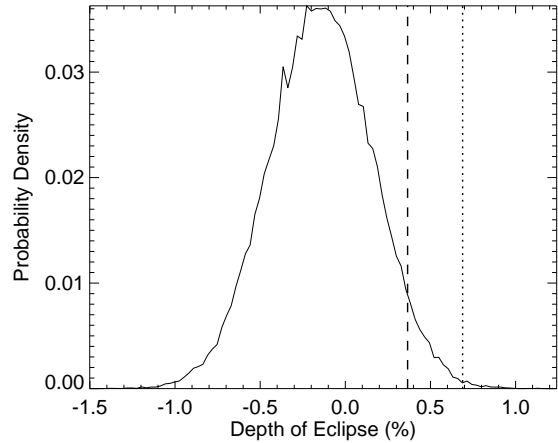}
\caption{Probability distribution for the eclipse depth from a Markov Chain Monte Carlo fit to the IRS data.  Negative values for the eclipse depth correspond to an increase in flux during the eclipse event, and positive values correspond to a decrease in flux (as would normally be expected for an eclipse).  The long-dashed and short-dashed lines indicate the $2\sigma$ and $3\sigma$ limits on the eclipse depth, respectively, which were calculated by integrating over this distribution.\label{hist}}
\end{figure}

Compared to this simple estimate, we find that the relative RMS variation in the time series of $3.3\%$ is a factor of 2.8 higher than expected.  This is surprising, but an examination of the scatter in the fluxes for individual background pixels across the images indicates that this background has a noise level three times higher than predicted based on photon and read noise alone.  This increased background noise could easily explain the increased scatter in the final time series.  \citet{dem06} observed a secondary eclipse of HD~189733b at 16~\micron~and found that they were able to obtain a RMS variation 1.7 times higher than the predicted photon noise, but this star is approximately 100 times brighter than TrES-4 in the infrared and the relative noise contribution from the sky background is correspondingly small in these observations.  Although the RMS variation in the background fluxes for our TrES-4 observation is inconsistent with the predicted photon and read noise alone, it \emph{is} consistent with the uncertainties produced by the Spitzer pipeline.  The \emph{Spitzer} error estimates include additional uncertainties from dark current subtraction, droop correction, flat fielding, and other other steps in the standard \emph{Spitzer} pipeline, therefore we speculate that it is one (or more) of these steps that is the dominant source of noise in our data.  

We place an upper limit on the eclipse depth in this bandpass by fitting the data simultaneously with a quadratic function of $ln(dt)$ as described in Eq. \ref{eq2} and an eclipse function where we have fixed the timing to the predicted value and allow the depth to vary over both positive and negative values.  We fit the data using a Markov chain Monte Carlo method as described in \S\ref{short_norm}, with uncertainties for individual points set equal to the RMS variation in the out-of-transit data after the detector ramp has been removed.  Because the star is not much brighter than the background, pixels at the center of the star's psf exhibit a similar ramp to that of the background pixels, and subtracting this background removes the majority of the detector ramp described in \S\ref{long_norm}.  There is still a small ramp remaining in the final time series (see upper panel of Figure \ref{irs_plot}), and we find that fitting this ramp with Eq. \ref{eq2} reduces the $\chi^2$ value of the best-fit solution by 0.5 when compared to a simple linear function of time.  Although this is not a large improvement, we note that the quadratic fit is a more accurate description of the detector ramp, and using it in our fits here allows to determine the upper limit on the eclipse depth in a manner consistent with our previous analysis in the 5.8 and 8.0~\micron~bandpasses. 

Our best-fit eclipse depth from these fits is $-0.14\pm0.30\%$, indicating that the measured flux increased by a statistically insignificant amount during the eclipse event.  Integrating over the probability distribution for this parameter from the Markov fit (see Fig. \ref{hist}), we find that the $2\sigma$ upper limit on the eclipse depth is $0.37\%$, and the $3\sigma$ upper limit is $0.69\%$.  We adopt the $2\sigma$ limit in the discussion below.

\begin{figure*}
\epsscale{0.85}
\plotone{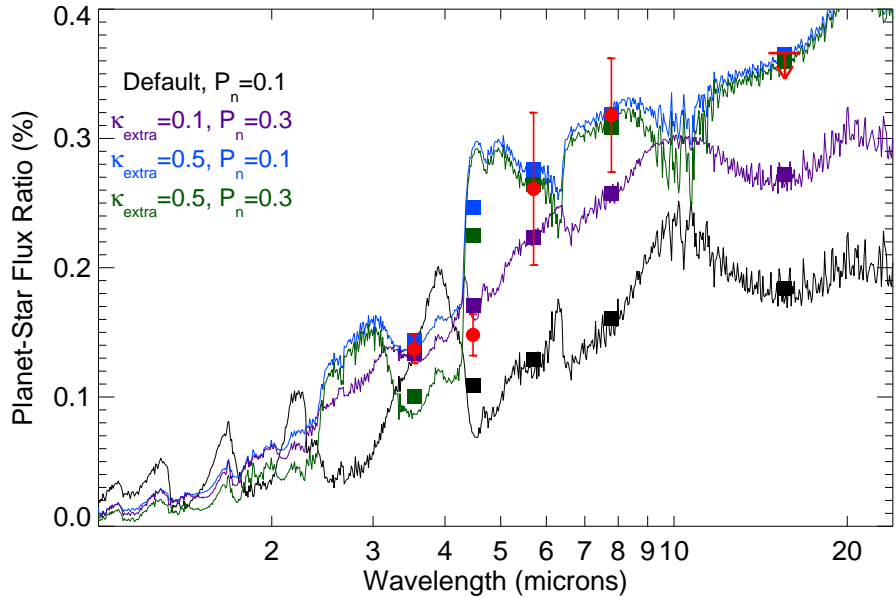}
\caption{Day-side planet-star flux ratios for TrES-4 as determined from measurements of the secondary eclipse depth in the four IRAC bandpasses (red circles).  The horizontal bar and arrow at 16~\micron~show the $2\sigma$~upper limit on the eclipse depth in this bandpass.  The black line corresponds to the default model (no temperature inversion) with a redistribution parameter $P_n=0.1$, which describes the case where 10\% of the incident energy is redistributed to the night side.  The purple, blue, and green lines correspond to models with an additional optical absorber at high altitudes (parameterized as $\kappa_{extra}$), which produces a thermal inversion around pressures of 0.001 bar \citep{burr07,burr08}.  Squares show the values for these models after integrating over the \emph{Spitzer} bandpasses.  The high planet-star flux ratios at 5.8 and 8.0~\micron~argue strongly for the presence of an inversion, as models with $\kappa_{extra}>0$ cm$^2$/g provide the best match at these wavelengths.  The eclipse depths in the 3.6 and 4.5~\micron~bandpasses are best matched by a model with relatively efficient day-night circulation and modest additional opacity (purple model, $\kappa_{extra}=0.1$ cm$^2$/g and $P_n=0.3$). \label{spectrum}}
\end{figure*}

\section{Discussion}

We find that the RMS variation in our final time series is 0.96, 1.06, 1.24, and 1.35 times the predicted photon noise from the star and background flux at 3.6, 4.5, 5.8, and 8.0~\micron, respectively.  In the 16~\micron~bandpass the noise is a factor of 2.8 higher than the predicted contribution from the photon noise of the star and background and the detector read noise, but it is consistent with the estimated uncertainties from the \emph{Spitzer} pipeline.  As a result of this increased uncertainty, we are unable to detect the secondary eclipse in the 16~\micron~bandpass, and instead place an upper limit on the eclipse depth at this wavelength.   

We determine the best-fit eclipse times for the two secondary eclipses observed using IRAC by taking the weighted average of the best-fit eclipse times in each bandpass.  Using this method, we find that the eclipse observed in the 3.6 and 5.8~\micron~bandpasses occurred $4.2\pm3.5$~minutes later than the predicted time based on the ephemeris from \citet{mand07}, where we have neglected the light travel time in the TrES-4 system \citep[on the order of 30 s;][]{loeb05} and assumed the secondary eclipse will occur exactly half an orbit after the transit.  We repeat this analysis for the eclipse observed in the 4.5 and 8.0~\micron~bandpasses, and find that it occurred $5.5\pm3.7$~minutes later than predicted.  If we assume that the planet's orbit remained the same over the 3.5-day period spanned by our observations (ie no perturbations that would change the orbital semi-major axis or eccentricity during this time), we can combine observations in all four of the IRAC bandpasses to get a single estimate for the best-fit eclipse time, which we find occurs $4.8\pm2.6$ minutes later than predicted.  However, there is an additional $\pm5.0$~minute uncertainty in the predicted transit time from \citet{mand07}.  Including this uncertainty we find the two averaged eclipses occur $4.8\pm5.6$ minutes later than predicted, which is consistent with zero offset.  

Our estimate for the best-fit timing offset translates to a constraint on the orbital eccentricity $e$ and the argument of pericenter $\omega$ of $ecos(\omega)=0.0015\pm0.0017$; the $3\sigma$ upper limit on this value is $|ecos(\omega)|<0.0058$, where we have calculated the limit by integrating over a Gaussian distribution with limits of integration that are symmetric around zero.  We selected these limits because we are interested in constraining the magnitude of $e$ rather than the sign of the $cos(\omega)$ term.  This upper limit means that unless the longitude of periastron $\omega$ is close to 90\degr~or 270\degr, we can rule out tidal heating from ongoing orbital circularization \citep{bod01,liu08} as an explanation for TrES-4's inflated radius.  \citet{liu08} estimate that TrES-4 would need to have an orbital eccentricity of approximately 0.04 to provide the required energy; this would require $|\omega|$ to be less than 9\degr~away from the two angles listed above to be consistent with our upper limit.  Our conclusion is also consistent with fits to the four radial velocity points from \citet{mand07}, although these points provide a relatively weak constraint on the eccentricity.

Next we compare the secondary eclipse depths in the four IRAC bandpasses to the predictions from atmosphere models for this planet (see Fig. \ref{spectrum} and \ref{ptplot} below). In order to fit the IRAC data and the 16-micron upper limit, we employed the same formalism described in \citet{burr07,burr08}.  Using the planet-star radius ratio of $0.09903\pm0.00088$ from \citet{mand07} and a Kurucz atmosphere model \citep{kurucz79,kurucz94,kurucz05} with an effective temperature of $6200\pm75$~K for the stellar spectrum \citep{sozz08}, we calculated the emergent spectrum at secondary eclipse for a pair of free parameters, P$_n$ and $\kappa_{extra}$.  P$_n$ is the dimensionless redistribution parameter that, in approximate fashion, accounts for the cooling of the dayside and the warming of the nightside by zonal winds near an optical depth of order unity.  It is a measure of the efficiency of heat redistribution by super-rotational hydrodynamic flows.  As the value of P$_n$ is increased, the day side becomes cooler and the emergent planetary flux at superior conjunction becomes correspondingly small.  $\kappa_{extra}$ is the absorptive opacity in the optical at altitude (here in cm$^2$/g) used to create a temperature inversion. The origin of such an absorber is unknown, but might be due to exotic species generated by the severe insolation regime or to TiO/VO at altitude \citep{hub03,burr07,burr08,fort08}.  Concerning the latter, the coldtrap effect should deplete the upper atmosphere of such diatomics, but atmospheric circulation and/or turbulence could alter this picture by advecting these compounds up from the lower atmosphere or removing them on the cooler night side of the planet.

\begin{figure}
\epsscale{1.2}
\plotone{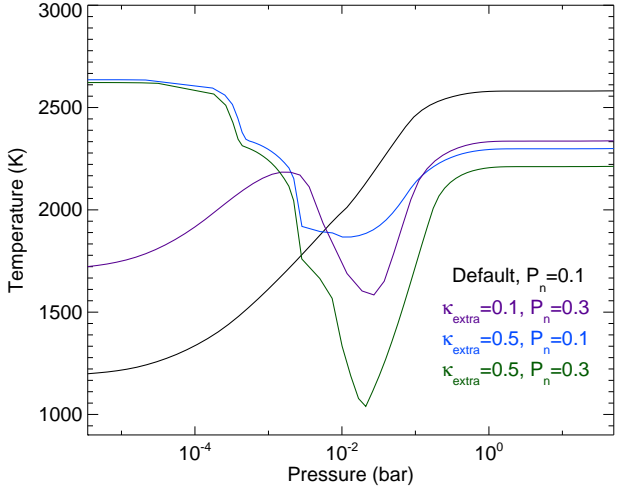}
\caption{Day-side pressure-temperature profiles for the four models plotted in Fig. \ref{spectrum}.  Increasing the opacity of the optical absorber (parameterized as $\kappa_{extra}$) increases the temperatures at pressures less than $0.01-0.001$ bars, while increasing the fraction of heat redistributed to the night side (parameterized as $P_n$) decreases the temperatures around pressures of 0.01-0.1 bars.  This is because \citet{burr08} parameterize the effects of energy transport to TrES-4's night side in these 1D models by adding a heat sink at a pressure of 0.1 bars, which causes a drop in temperature as the day-night circulation is turned up.  See \citet{burr07,burr08} for a full description of these parameterizations and the corresponding models.\label{ptplot}}
\end{figure}

Fig. \ref{spectrum} and \ref{ptplot} show four models with varying values for $P_n$ and $\kappa_{extra}$.  The standard non-inverted model ($\kappa_{extra}=0$ cm$^2$/g) is clearly inconsistent with the observed fluxes from TrES-4 at wavelengths longer than 4~\micron.  It is possible to match the observed 3.6~\micron~flux with this model by reducing the relative fraction of the incident energy that is redistributed to the planet's night side, thus increasing the day-side temperature and corresponding fluxes, but even this change is insufficient at longer wavelengths.  In contrast to this model, all three models with a thermal inversion ($\kappa_{extra}>0$ cm$^2$/g) provide an improved match to the 5.6 and 8.0~\micron~fluxes.  The ratio of the 3.6~\micron~and 4.5~\micron~fluxes, another measure of inversion (Burrows et al. 2007), is also less than one, much lower than for models without inversions.  The best overall fit is obtained by setting $P_n=0.3$ and $\kappa_{extra}=0.1$ cm$^2$/g, corresponding to a case with relatively efficient day-night circulation and modest additional opacity.  

We note that our particular choice of planet-star radius radio and stellar effective temperature may affect the predictions of these models.  As a test we re-run our full radiative transfer codes for the $P_n=0.3$ and $\kappa_{extra}=0.1$ cm$^2$/g model using stellar atmosphere models with temperatures of 6100~K and 6300~K, and measure the resulting change in the predicted eclipse depths in the four IRAC channels.  We find that decreasing the stellar effective temperature by 100~K shifts the predicted eclipse depths in the [$3.6$,$4.5$,$5.8$,$8.0$]~\micron~bandpasses by [$-1.8\%$,$-1.3\%$,$-1.1\%$,$-0.9\%$], while increasing the temperature by 100~K results in changes of [$+1.7\%$,$+1.4\%$,$+1.1\%$,$+0.9\%$] in these same bandpasses.  These changes are negligible relative to the $1\sigma$ uncertainties of [$8.0\%$,$10.8\%$,$22.6\%$,$13.8\%$] in the measured eclipse depths.  From this test we conclude that increasing or decreasing the effective temperature of the star by an amount comparable to the formal uncertainties in this quantity cannot produce a non-inverted model that is consistent with the measured eclipse depths at longer wavelengths, and is unlikely to alter our conclusion that the $P_n=0.3$ and $\kappa_{extra}=0.1$ cm$^2$/g model provides the best fit to the data.  The predicted eclipse depths scale linearly with the planet-star area ratio, but squaring the planet-star radius ratio from \citet{mand07} results in a value of $0.98\pm0.02\%$ for this quantity; this is a factor of 100 smaller than the typical uncertainty contributed by the stellar effective temperature.

It is interesting to note that the same values of $\kappa_{extra}=0.1$ cm$^2$/g and $P_n=0.3$ also provide the best fit to the observed broadband emission spectra for HD 209458b \citep{burr07,burr08}.  This would seem to imply that the atmospheric circulation and relative abundances of the species responsible for creating the inversions in the upper atmospheres of both planets are similar, despite the higher temperatures and increased levels of irradiation experienced by TrES-4.  This is probably an oversimplification of the problem, however, as there are likely substantial thermal and chemical gradients between the substellar point and the day-night terminator on both planets, and our observations constrain only the hemisphere-averaged properties of the day-side atmosphere.     

\section{Conclusions}

Our observations of TrES-4 at 3.6, 4.5, 5.8, 8.0, and 16.0~\micron~reveal that this planet has a thermal inversion similar to the one observed for HD 209458b \citep{knut08,burr07}.  The presence of an inversion in the atmosphere of TrES-4 provides support for the idea that planets with higher levels of irradiation are more likely to have thermal inversions, although it does not distinguish between competing theories for the nature of the optical absorber responsible for the creation of the inversions.  

If we are to fully understand the mysterious origin of these temperature inversions, it will require a much larger sample than the seven systems with published \emph{Spitzer} observations.  There are currently 40 known transiting planetary systems with published coordinates, of which 33 are bright enough to observe with \emph{Spitzer}.  A sample spanning a range of stellar metallicities, levels of irradiation, surface gravities, and orbital periods might reveal important connections between the presence of a temperature inversion and other properties of the system.  Such a comprehensive survey has the potential to provide an explanation for XO-1b's thermal inversion \citep{mach08}, which is difficult to understand within the current irradiation-dependent picture.  

We estimate an upper limit of $|ecos(\omega)|<0.0058$ for the orbital eccentricity, consistent with the fits to the radial velocity data presented by \citet{mand07}.  This upper limit means that unless the longitude of periastron $\omega$ is close to 90\degr or 270\degr, we can rule out tidal heating from ongoing orbital circularization at the level required by \citet{liu08} in order to explain TrES-4's inflated radius.

Although \emph{Spitzer} will deplete its supply of cryogen in spring 2009, the 3.6 and 4.5~\micron~channels will continue to function at full sensitivity, and a two-year mission using these two channels was recently approved by NASA. Observations of the secondary eclipse in these two channels should be sufficient to distinguish between planets with and without temperature inversions, as demonstrated by our discussions above.  These observations can be compared to the set of more than a dozen planets for which there will be secondary eclipse observations in all four IRAC channels by the end of the cryogenic \emph{Spitzer} mission.  This basis set of comprehensive observations, when combined with a significantly larger survey in the 3.6 and 4.5~\micron~channels, should provide the statistical leverage needed to resolve the origin and nature of these inversions. 

\acknowledgements

This work is based on observations made with the \emph{Spitzer Space Telescope}, which is operated by the Jet Propulsion Laboratory, California Institute of Technology, under contract to NASA.  Support for this work was provided by NASA through an award issued by JPL/Caltech.  HAK was supported by a National Science Foundation Graduate Research Fellowship.  AB would like to acknowledge the support of NASA through grant NNX07AG80G and JPL/Spitzer Agreement No. 1328902 and through the NASA Astrobiology Institute under Cooperative Agreement No. CAN-02-OSS-02 issued through the Office of Space Science.  F. T. O'D. was supported by an appointment to the NASA Postdoctoral Program at the Goddard Space Flight Center, administered by Oak Ridge Associated Universities through a contract with NASA.

\end{document}